\documentclass[cameraready]{Interspeech}

\title{CAST-TTS: A Simple Cross-Attention Framework for Unified Timbre Control in TTS}

\author[affiliation={1,2}]{Zihao}{Zheng}
\author[affiliation={1}]{Wen}{Wu}
\author[affiliation={1}]{Chao}{Zhang}
\author[affiliation={2}]{Mengyue}{Wu}
\author[affiliation={1}]{Xuenan}{Xu}

\address{
    $^1$ Shanghai AI Lab, China \\
    $^2$ MoE Key Lab of Artificial Intelligence, X-LANCE Lab, Shanghai Jiao Tong University, China
}

\email{rookie9@sjtu.edu.cn, wsntxxn@gmail.com}

\keywords{text-to-speech synthesis, Timbre Control, computational paralinguistics}

\usepackage{comment}
\usepackage{amsmath}
\usepackage{mathtools}
\usepackage{booktabs} 
\usepackage{multirow}
\usepackage{cleveref}
\usepackage{hyperref}
\usepackage{xcolor}
\usepackage{xcolor}

\newcommand\blfootnote[1]{%
  \begingroup
  \renewcommand\thefootnote{}\footnote{#1}%
  \addtocounter{footnote}{-1}%
  \endgroup
}

\begin{document}

\maketitle

\begin{abstract}
    Current Text-to-Speech (TTS) systems typically use separate models for speech-prompted and text-prompted timbre control. While unifying both control signals into a single model is desirable, the challenge of cross-modal alignment often results in overly complex architectures and training objective.
    To address this challenge, we propose CAST-TTS, a simple yet effective framework for unified timbre control.
    Features are extracted from speech prompts and text prompts using pre-trained encoders.
    The multi-stage training strategy efficiently aligns the speech and projected text representations within a shared embedding space.
    A single cross-attention mechanism then allows the model to use either of these representations to control the timbre.
    Extensive experiments validate that the unified cross-attention mechanism is critical for achieving high-quality synthesis.
    CAST-TTS achieves performance comparable to specialized single-input models while operating within a unified architecture.
    The demo page can be accessed at \href{https://HiRookie9.github.io/CAST-TTS-Page}{\textcolor{cyan}{\textit{https://HiRookie9.github.io/CAST-TTS-Page}}}
\end{abstract}
\blfootnote{Under review at Interspeech 2026.}
\section{Introduction}

In recent years, Text-to-Speech (TTS) systems have made significant progress, enabling the generation of speech with high naturalness and fidelity.
Recent TTS models often possess zero-shot voice cloning capabilities, allowing them to mimic a speaker's timbre from a short speech prompt~\cite{chen2025f5, eskimez2024e2, wang2024maskgct, zhu2025zipvoice}.
By training on a speech-infilling task, zero-shot TTS models based on flow-matching have shown strong performance ~\cite{chen2025f5,eskimez2024e2, le2023voicebox}.


A parallel line of research explores the use of text prompts to control speaker characteristics~\cite{lacombe-etal-2024-parler-tts, wang2025capspeech}.
In contrast to speech prompts, textual captions provide a more coarse-grained but flexible approach to controlling speaker timbre.
Large language models (LLMs) and expert systems are involved in automatic data annotation pipelines, facilitating the construction of large-scale, open-source datasets~\cite{lyth2024natural}.
In representative works such as Parler-TTS~\cite{lacombe-etal-2024-parler-tts} and CapSpeech~\cite{wang2025capspeech}, textual captions are first encoded by a pre-trained text encoder.
The resulting embeddings then condition the synthesis model through a cross-attention mechanism, which has proven effective for accurate timbre attribute control.
The effectiveness of this conditioning mechanism has also been demonstrated in Text-to-Image (T2I) and Text-to-Audio (T2A) models~\cite{rombach2022high, hai2025ezaudio}.

Models that exclusively use either a \textit{speech prompt} or a \textit{text prompt} often fail to meet the diverse needs of users across different scenarios.
Consequently, prior research has begun to explore the combination of these two forms.
StyleFusion TTS~\cite{chen2024stylefusion} integrates style descriptions with audio prompts, but it requires both inputs simultaneously during inference.
FleSpeech offers a more flexible solution, enabling various combinations of audio, text, and even facial inputs to enhance adaptability~\cite{li2025flespeech}.
However, FleSpeech employs a complex architecture comprising an autoregressive language model, a flow-matching backbone, and a diffusion-based prompt encoder.
The design necessitates optimizing several loss functions simultaneously for semantic prediction, feature reconstruction, and prompt alignment, thereby increasing the risk of training instability.

In this work, we introduce \textbf{CAST-TTS}, a framework empolying \textbf{C}ross-\textbf{A}ttention mechanism combining either \textbf{S}peech or \textbf{T}ext prompts.
For text prompts, we use Flan-T5~\cite{chung2024scaling} to encode the descriptive text.
The resulting features are then projected into the shared timbre embedding space via a lightweight projector.
For speech prompts, we follow the procedure of E2-TTS-x1~\cite{eskimez2024e2}. 
An original speech utterance is randomly split into prompt and target.
Montreal Forced Aligner (MFA) is used to obtain transcriptions for the target part~\cite{mcauliffe2017montreal}.
For the speaker timbre encoding, we employ a pre-trained speaker encoder~\cite{desplanques2020ecapa} to extract embeddings from the prompt part. 
This design eliminates the masking strategy in E2-TTS and results in a consistent conditioning paradigm of the timbre under speech and text prompt scenarios.


\begin{figure*}[ht]
    \centering
    \includegraphics[width=0.95\textwidth]{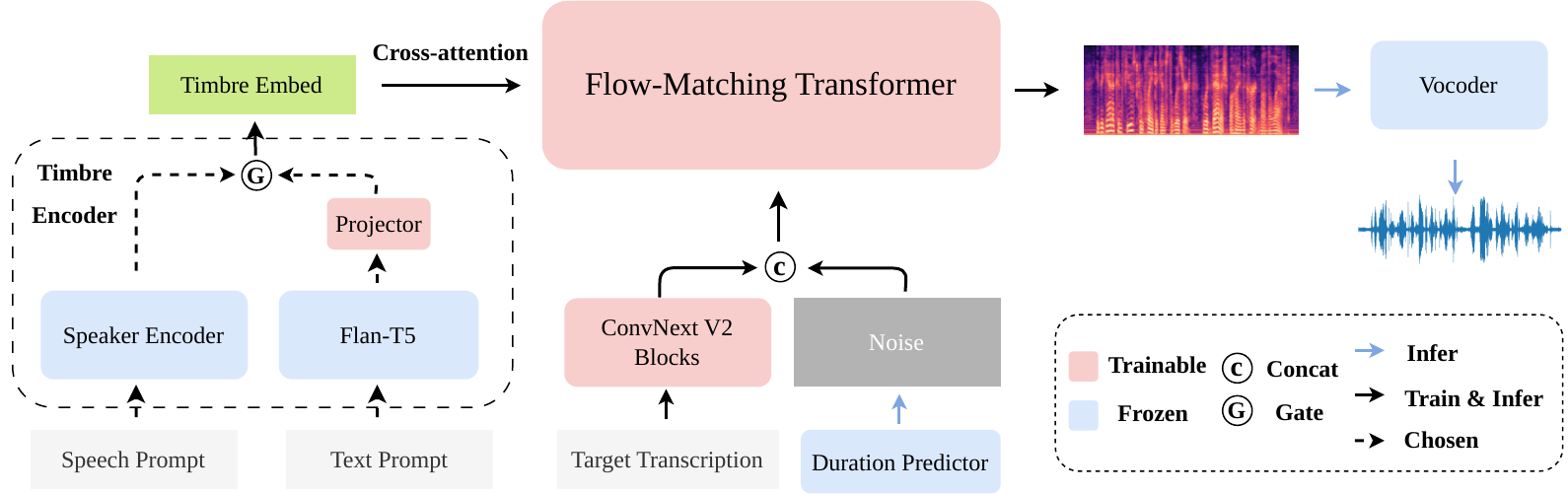}
    \caption{An overview of CAST-TTS. The timbre encoder converts speech or text prompts into timbre embeddings, which condition the synthesis model through a cross-attention mechanism.}
    \label{fig:model}
\end{figure*}

Following established practices, the target text transcription is first encoded by a ConvNeXt V2 module, then concatenated with noisy latent~\cite{woo2023convnext}. Separately, speaker information from either modality is incorporated into the model via cross-attention.
We adopt a multi-stage training strategy to optimize and align the shared timbre embedding space.
In summary, our contributions are as follows:
\begin{itemize}
    \item A simple architecture that uses cross-attention to fuse speech prompt conditions. This removes the requirement for complex masking and is aligned with text prompts conditioning.
    \item A unified framework capable of handling both speech and textual modalities by projecting text embeddings into speech-based timbre embedding space with a multi-stage training strategy.
    \item Extensive ablation studies and comparative results that validate our design choices and demonstrate the effectiveness of the unified model.
\end{itemize}


\section{CAST-TTS}
CAST-TTS is a non-autoregressive (NAR) TTS system designed to integrate speech-prompted and text-prompted timbre control within a single framework.
To achieve this, a unified timbre encoder is employed to encode the inputs of both modalities into a shared embedding space.
We further design a multi-stage training strategy to optimize the cross-modal feature alignment.
\subsection{Architecture}
As illustrated in \Cref{fig:model}, CAST-TTS mainly consists of a timbre encoder and a Transformer flow-matching backbone.
The timbre encoder processes either a speech prompt or a text prompt, to generate an embedding that controls speaker timbre.
The Transformer backbone predicts the target mel-spectrogram, which is subsequently converted into an audio waveform by a BigVGAN vocoder~\cite{lee2023bigvgan}.

The timbre encoder comprises two parts: the speech branch and the text branch. 
The speech branch is a speaker encoder that transforms the input speech prompt into a timbre embedding sequence $\mathbf{T} \in \mathbb{R}^{T\times D}$.
Based on a comparison of different models in \Cref{sec:speakerfeatures}, we select the encoder of a WavLM-based~\cite{chen2022wavlm} ECAPA-TDNN ~\cite{desplanques2020ecapa}.
In the text branch, the input caption is encoded by a Flan-T5 encoder~\cite{chung2024scaling}.
A linear projector then maps the text embeddings into the timbre embedding space, since speech prompts inherently contain richer and more fine-grained speaker information than textual descriptions.
This directional alignment designates the rich speech embedding space as the unified conditioning space, forcing the text features to align with the more expressive speech modality.

The input text transcription is first converted into a sequence of character embeddings, then padded with filler tokens to match the length of the target mel-spectrogram.
We employ ConvNeXt V2 blocks for character encoding, given the proven ability in prior works~\cite{chen2025f5}.

The Transformer backbone incorporates several Transformer blocks with long skip connections between blocks.
Zero-initialized adaptive Layer Norm (adaLN-zero) is used to stabilize training.
The Transformer backbone receives three inputs: the noisy mel-spectrogram $\mathbf{M}$, character embeddings $\mathbf{C}$, and timbre embeddings $\mathbf{T}$.
First, the noisy mel-spectrogram is concatenated with the character embeddings. Within each block of the Transformer, the latent representation is encoded through self-attention.
Subsequently, it interacts with the timbre embedding via cross-attention, followed by a final feed-forward network (FFN).

\subsection{Training}
CAST-TTS is trained with flow-matching loss. This objective trains a neural network $v_\theta$ to parameterize a velocity field that defines a continuous-time flow between the data distribution $\mathbf{x}_0 \sim p_\text{data}(x_0)$ and a standard Gaussian prior $\mathbf{x}_1 \sim \mathcal{N}(0, \mathbf{I})$:
\begin{align*}
    \frac{d \mathbf{x}_\tau}{d\tau} &= v_\theta(\mathbf{x}_\tau, \tau)\\
    \mathbf{x}_\tau &= (1-\tau)\cdot\mathbf{x}_0 + \tau\cdot\mathbf{x}_1, \quad \tau \in [0,1] \\
    \mathcal{L}_{\mathrm{FM}} &= \mathbb{E}_{\tau, \mathbf{x}_0, \mathbf{x}_1}  \left\lVert v_\theta(\mathbf{x}_\tau, \tau, \mathbf{C}, \mathbf{T}) - (\mathbf{x}_1 - \mathbf{x}_0) \right\lVert^2
\end{align*}
where $\theta$ denotes model parameters, $\tau$ is the flow step, and $\mathcal{L}_{\mathrm{FM}}$ is the flow-matching training loss.
Throughout the training process, the pre-trained speaker encoder and the Flan-T5 text encoder remain frozen.
\begin{table*}[t]
\centering
\caption{Objective and subjective evaluation results. The arrows (↓↑) indicate whether lower or higher scores are better.}
\label{tab:main_results}
\begin{tabular}{ll|cccc|cc} 
\toprule
\multicolumn{2}{c}{} & \multicolumn{4}{c}{\textbf{Objective Evaluations}} & \multicolumn{2}{c}{\textbf{Subjective Evaluations}} \\
\midrule
\textbf{Prompt} & \textbf{Model} & \textbf{WER(\%)↓} & \textbf{SPK-Sim↑} & \textbf{Style-ACC(\%)↑} & \textbf{UTMOS↑} & \textbf{N-MOS↑} & \textbf{Sim-MOS↑} \\ 
\midrule
\multirow{4}{*}{Speech} & F5-TTS-v1 & 2.31&75.4& -& 3.87& \textbf{4.13}& \textbf{4.17}\\
 & MaskGCT & 3.54& 74.5& -& 3.90& 3.73& 3.93\\
 & ZipVoice-L & \textbf{1.77}& 66.7&- & \textbf{4.26}& 3.90&4.01 \\
 & \textbf{CAST-TTS} & 2.05&\textbf{78.4} &- &3.91&3.86 &4.09 \\ 
\midrule
\multirow{3}{*}{Text} & CapSpeech-NAR& 5.11&- & 88.93& \textbf{4.06}& \textbf{4.08}& 4.05\\
 & Parler-TTS-Large& 5.53&- & 82.04& 3.80& 3.46& 3.52\\
 & \textbf{CAST-TTS} &\textbf{3.89} &- & \textbf{91.15}& 4.01& 4.03& \textbf{4.11}\\ 
\bottomrule
\end{tabular}
\end{table*}

CAST-TTS is trained in three distinct stages. 
This process utilizes two distinct data formats: speech-prompted pairs and text-prompted pairs, which are detailed further in \Cref{sec:data_pipeline}:
\begin{itemize}
    \item 
\textbf{Speech Synthesis Pre-training:} We first train the ConvNeXt V2 blocks and Transformer layers using only the speech-prompted dataset. 
This stage equips the model with a basic ability to generate coherent speech conditioned on timbre embeddings.
    \item 
\textbf{Text Condition Alignment:} Next, we freeze the pre-trained components and train \textbf{only the projector} on the text-prompted dataset. 
This efficiently aligns the text representation space with the established speech-derived timbre embedding space.
    \item \textbf{Joint Fine-tuning:} 
Finally, all trainable components are jointly fine-tuned on the combined dataset. 
This stage refines the alignment between both modalities and enhances overall synthesis quality and controllability.
\end{itemize}

\subsection{Inference}
CAST-TTS generates target speech based on a target transcription $T_{gen}$ and a speech or text prompt.
Since NAR models cannot inherently predict audio duration, an external duration predictor is required to determine the length of the output speech.
For speech prompt, we employ Whisper-large-v3 extracting the reference transcription $T_{ref}$~\cite{radford2023robust}.
Following previous works, we estimate target duration based on the ratio of character counts in $T_{ref}$ and $T_{gen}$. 
For text prompt, we leverage the pre-trained duration predictor from CapSpeech, which accepts both the transcription and the prompt to estimate the total duration of the target speech~\cite{wang2025capspeech}.
During inference, classifier-free guidance (CFG) is employed to improve the generation quality~\cite{ho2021classifier}:
\begin{align*}
v_\theta^{\text{CFG}}(\mathbf{x}_\tau, \mathbf{C}, \mathbf{T}) 
= (1-w)v_\theta(\mathbf{x}_\tau, \varnothing, \varnothing) 
+ w v_\theta(\mathbf{x}_\tau, \mathbf{C}, \mathbf{T}) 
\end{align*}
where $w$ is the guidance scale.
\section{Experiments}

\subsection{Datasets}
\label{sec:data_pipeline}
Since CAST-TTS synthesizes speech conditioned on speech or text prompts, the training data comprises two corresponding types of data pairs.

LibriTTS-R dataset~\cite{koizumi2023libritts} is selected for speech-prompted training. We obtain word-level alignments for each utterance using the MFA model, following the methodology of E2-TTS-x1~\cite{eskimez2024e2}.
For each sample, we determine a random split timestamp.
The audio segment preceding this timestamp serves as the speech prompt.
The remaining audio segment and its corresponding transcription become the target and input respectively.

For natural language conditioning, we mainly use the LibriTTS-R subset of CapTTS dataset~\cite{wang2025capspeech}.
A part of GigaSpeech is included to add child, teen, and elderly speakers~\cite{chen2021gigaspeech}, since LibriTTS-R lacks speakers of these ages. 
CapTTS employs specialized models to annotate discrete labels for speaker attributes: gender, accent, pitch, tonal expressiveness and speaking rate.
Subsequently, LLM is used to generate descriptive captions from these labels.


In total, the training data is comprised of approximately 1360 hours of audio, including $\sim$282K speech-prompted data pairs and $\sim$434K text-prompted data pairs.
\subsection{Experimental Setup}
The Transformer backbone follows CapSpeech, with 22 layers, 16 attention heads and a hidden size of 1024.
The training details in three stages are:
\begin{itemize}
    \item Stage 1: 400K steps with a peak learning rate of 7.5e-5.
    \item Stage 2: 200K steps with a peak learning rate of 1.5e-5.
    \item Stage 3: 100K steps with a peak learning rate of 2.5e-5.
\end{itemize}
The learning rate takes a linear decay schedule with a weight decay
of 0.01. During inference, the CFG scale is
set to 3.0.

\subsection{Evaluation}
We evaluate the model's performance on two distinct test sets, each targeting a different conditioning modality. For speech-prompted evaluation, we choose LibriSpeech-PC \textit{test-clean}~\cite{meister2023librispeech}. For text-prompted evaluation, we use the CapTTS test subsets corresponding to the LibriTTS-R and GigaSpeech audio.

Objective metrics include Word Error Rate (WER), Speaker Similarity (SPK-Sim), Style-ACC and UTMOS. WER is computed using whisper-large-v3 along with the Whisper text normalizer.
TitaNet~\cite{koluguri2022titanet} is used for SPK-Sim evaluation, computing the similarity of speaker embeddings between prompt and generated speech.
Style-ACC is the average accuracy of age, gender, pitch, expressiveness of tone, and speed.
The reliance on hard thresholds for classifying continuous predictions makes the baseline metrics for pitch, expressiveness, and speed sensitive to minor value changes near the boundaries.
We therefore relax the evaluation criterion by allowing predictions to fall within adjacent categories, which is more stable and representative.
UTMOS is used for evaluating audio quality~\cite{saeki2022utmos}.

Subjective metrics include the Mean Opinion Score (MOS) to evaluate speech naturalness (N-MOS) and similarity (Sim-MOS).
10 highly educated raters without hearing loss are invited to score 10 random samples from each of the test dataset.

\section{Results}
\subsection{Generation Performance}
For the speech-prompted synthesis task, we benchmark our CAST-TTS against several leading models: F5-TTS-v1~\cite{chen2025f5}, MaskGCT~\cite{wang2024maskgct}, and the LibriTTS-trained version of ZipVoice (denoted as ZipVoice-L)~\cite{zhu2025zipvoice}.
The results in \Cref{tab:main_results} demonstrate that CAST-TTS achieves the highest SPK-SIM among all models and delivers competitive WER and UTMOS scores.
F5-TTS-v1 demonstrates a clear advantage in the subjective ratings, likely due to its training on the large-scale Emilia dataset.
Notably, CAST-TTS achieves performance on par with the other competing TTS systems.
This indicates that CAST-TTS possesses strong synthesis performance and excellent speaker-cloning capabilities.
\begin{table*}[t]
\centering
\caption{Ablation study on model architectures. We evaluate the impact of different methods under both speech and text prompted conditions.}
\label{tab:model_ablation}
\begin{tabular}{ll|ccc|ccc}
\toprule
\multicolumn{2}{c}{} &\multicolumn{3}{c}{\textbf{Speech Prompt}} & \multicolumn{3}{c}{\textbf{Text Prompt}} \\
\textbf{Model } &\textbf{Feature} & \textbf{WER(\%)↓} & \textbf{SPK-Sim↑} & \textbf{UTMOS↑} & \textbf{WER(\%)↓} & \textbf{Style-ACC(\%)↑} & \textbf{UTMOS↑} \\
\midrule
\multirow{2}{*}{CAST-SA}      &  Melspec& 3.74& 35.8& 3.97& \textbf{4.17}& 81.25&3.78 \\
&ECAPA-TDNN  &5.48 &43.0 &4.00 & 6.56& 90.00&\textbf{3.91} \\
\midrule
\multirow{2}{*}{CAST-SACA}            &Melspec & \textbf{2.67}& 35.5& 3.95& 4.52& 90.10& 3.89 \\
 &ECAPA-TDNN& 3.18&41.2& \textbf{4.06}&4.93 &\textbf{90.26} & 3.90\\
\midrule
CAST-CA            &  ECAPA-TDNN & 3.13& \textbf{69.5}&3.82 & 4.47& 89.01&\textbf{3.91}\\
\midrule
CAST-TTS-BASE            &  \multirow{3}{*}{ECAPA-TDNN}&2.87 &75.4& 3.87& 4.03& 87.96& 3.85\\
CAST-TTS-TV            &  & 2.85& 77.2& 3.79& 4.08& 87.98& 3.81\\
CAST-TTS            &  & \textbf{2.05}& \textbf{78.4}& \textbf{3.91}& \textbf{3.89}& \textbf{91.15}& \textbf{4.01}\\
\bottomrule
\end{tabular}
\end{table*}

In the text-prompted synthesis task, CAST-TTS is compared against CapSpeech-NAR~\cite{wang2025capspeech} and Parler-TTS-Large ~\cite{lacombe-etal-2024-parler-tts}.
CAST-TTS achieves the best results in both WER and Style-ACC, while also yielding a strong UTMOS score.
For subjective metrics, CAST-TTS also exhibits highly competitive performance.
These results highlight the superior performance of CAST-TTS in generating high-fidelity speech that accurately reflects the guidance of text-based style captions.

\subsection{Ablation Study}
To validate our method, we conduct ablation studies investigating the effects of the speaker feature, the fusion mechanism, and the training strategy.
\subsubsection{Speaker Features}
\label{sec:speakerfeatures}
To investigate the effectiveness of feature representation for the speech prompt, we compare three different features, all integrated into the backbone via the cross-attention mechanism.
Mel-spectrogram is usually fused with the noisy latent by concatenation followed by self-attention, rather than cross-attention in prior works.
ECAPA-TDNN and TitaNet features are extracted from pre-trained speaker verification models. 
For a fair comparison of SPK-Sim, we compute scores using TitaNet and ECAPA-TDNN respectively, denoted as Sim-T and Sim-E.
\begin{table}[h]
\centering
\caption{Ablation results for speech features.}
\label{tab:feature_ablation}
\begin{tabular}{lccc}
\toprule
\textbf{Feature} & \textbf{WER(\%)↓} & \textbf{Sim-T↑} & \textbf{Sim-E↑}  \\
\midrule
Melspec       &3.41	&	47.9&	32.8               \\
TitaNet        &3.50		&\textbf{80.9}	&64.4               \\
ECAPA-TDNN        &\textbf{2.51}		&80.0	&\textbf{72.8}                \\
\bottomrule
\end{tabular}
\end{table}

As shown in \Cref{tab:feature_ablation}, the mel-spectrogram yields a substantially lower SPK-SIM score compared to both dedicated speaker features, since mel-spectrogram contains a mixture of acoustic and semantic information.
In contrast, speaker features are explicitly trained to disentangle and represent only the speaker-relevant acoustic characteristics, which allows the model to learn the target speaker identity more directly and effectively from the speech prompt.
Since ECAPA-TDNN features demonstrate more robust overall performance across the metrics than TitaNet features, we select the encoder of ECAPA-TDNN as the speaker encoder for CAST-TTS.
\subsubsection{Fusion Mechanism}
We conduct an ablation study to investigate the optimal mechanism for fusing the speech and text prompt embeddings.
We design and compare three distinct architectures:
\begin{itemize}
    \item CAST-SA: This approach concatenates the timbre embedding with the noisy latent, relying on self-attention for fusion. We test it with both mel-spectrogram and ECAPA-TDNN features, since mel-spectrogram is a common choice for this fusion style in prior works~\cite{chen2025f5, eskimez2024e2}.
    \item CAST-SACA: This model represents an intuitive hybrid approach. It fuses the speech prompt via concatenation and self-attention, while integrating the text prompt embedding using cross-attention.
    \item CAST-CA: This is our proposed architecture, which uses cross-attention to uniformly fuse both the speech and text prompt representation.
\end{itemize}
All models are trained for 400K steps on the full dataset for a fair comparison.
The results are shown in \Cref{tab:model_ablation}.
Overall, CAST-SACA models perform better than CAST-SA.
Regarding features, while mel-spectrogram yields a lower WER, ECAPA-TDNN is superior across nearly all other indicators.
CAST-CA achieves highly competitive WER, Style-ACC, and UTMOS scores while demonstrating a clear advantage in SPK-SIM.
This validates the effectiveness of using cross-attention as the primary mechanism for integrating timbre features.

\subsubsection{Training Strategy}
We investigate three training strategies to validate our proposed approach.
First, we train CAST-TTS-CA directly in an end-to-end fashion as a baseline (CAST-TTS-BASE).
Second, we evaluate CAST-TTS-TV, which prepends a learnable task vector to the timbre embedding to explicitly differentiate between modality types. 
The third method is our proposed multi-stage training strategy.
To ensure a fair comparison, all models are trained for a total of 700K steps.

As shown in \Cref{tab:model_ablation}, CAST-TTS-TV exhibits no improved performance over the CAST-TTS-BASE baseline after incorporating a task vector.
This result suggests that explicit distinction between the two tasks does not yield better performance.
The superior performance of CAST-TTS across all metrics highlights the advantages of our training strategy.
This approach allows the model to first establish a robust speech synthesis foundation and then efficiently align the text modality to the pre-trained space before a final joint refinement.

\section{Conclusion}
A unified model supporting both speech and text prompts offers significant flexibility for TTS, yet existing solutions are often architecturally complex due to the inherent difficulty of cross-modal alignment.
To address this issue, we propose CAST-TTS: a simple cross-attention framework for unified timbre control. 
We extract features from the input speech or text prompt using pre-trained encoders, mapping the text embedding into a unified timbre embedding space with a simple projector. 
To fuse these control signals, we employ cross-attention and propose a multi-stage training strategy for optimized cross-modal alignment. 
CAST-TTS achieves strong results on both speech-prompted and text-prompted synthesis tasks. 
Due to limitations in our dataset, it currently lacks control over attributes such as emotion and accent, which we identify as important directions for our future work.

\section{Generative AI Use Disclosure}
In this work, generative AI was exclusively utilized to fix grammatical mistakes and adjust terminology, while all core research activities, including study design, data collection, analysis, and scientific reasoning, were conducted independently by the authors.
\bibliographystyle{IEEEtran}
\bibliography{mybib}

\end{document}